# The BeiHang Keystroke Dynamics Authentication System

Juan Liu[1], Baochang Zhang[1], Linlin Shen[2], Jianzhuang Liu[3,4], Jason Zhao[5]

[1] *Science and Technology on Aircraft Control Laboratory, Beihang University, Beijing, China*

[2] *School of Computer Science and Software Engineering, Shenzhen University, China*

[3] *Shenzhen Key Lab for CVPR, Shenzhen Institutes of Advanced Technology, Chinese Academy of Sciences, China*

[4] *Department of Information Engineering, The Chinese University of Hong Kong, China*

[2] *Microdone Network Technologies Co., Ltd, China*

Email: bczhang@buaa.edu.cn, llshen@szu.edu.cn, jz.liu@siat.ac.cn

## Abstract

Keystroke Dynamics is an important biometric solution for person authentication. Based upon keystroke dynamics, this paper designs an embedded password protection device, develops an online system, collects two public databases for promoting the research on keystroke authentication, exploits the Gabor filter bank to characterize keystroke dynamics, and provides benchmark results of three popular classification algorithms, one-class support vector machine, Gaussian classifier, and nearest neighbour classifier.

## 1. Introduction

Electronic Devices with internet access are an essential part of modern society. Especially, Internet has greatly changed our lives, making our work and daily lives more convenient. However, it also brings us a big concern in information security. We depend so much on computers and Internet to store and process sensitive data, and it has become extremely necessary to secure them from intruders [1-5]. Our private information suffers from more risks than ever before.

To protect our information, we need to verify whether a user is legal using authentication and identification techniques. For user authentication and identification in computer based applications, there is a need for simple, low-cost and unobtrusive device. A user can be defined as a person who attempts to access information on the computer or internet in using the keyboard or touch screen. Currently, password is extensively and widely used to prevent user accounts from being intruded. However, too many methods can be used to decipher password, and once it is cracked, it will probably lead to a significant financial loss of the user would be caused. Such crimes on the Internet can cause a wide range of serious damages, and yet are difficult to be prevented. Therefore, to cope with such problems, we urgently need a more reliable way to protect our privacy.

Though the use of biometrics such as face, fingerprint and signature can improve the security, such techniques usually require additional tools to protect a device, which induces an extra cost in comparison with the password technique. The use of keystroke dynamics, which detects the typing pattern of an individual, can be an alternative for enhancing security without extra cost, as it can be obtained using the existing systems such as the standard keyboard. Keystroke dynamics is a kind of behavior feature [1,2]. It utilizes the rhythm and manner in which an individual types characters on a keyboard. The keystroke data contain the time of the key press and release (shown in Fig.1) from which two kinds of features are extracted, *flight time* and *dwelling time*. The flight time is defined as the time difference between one key release and the following key press. The dwelling time is the time difference between the press and release of one key. One of the major advantages of this biometric is that it is non-intrusive and can be applied covertly to enhance existing cyber-security systems. It is also suitable to the touch screen of current mobile devices [1-5], which also contains two kinds of features: *flight time* and *dwelling time*. However, physical keyboards still remain the primary way for data entry in many electronic devices. Hence, in this paper we focus on keystroke dynamics on physical keyboards.

In 1980s, the National Science Foundation and the National Bureau of Standards in the United States conducted some studies concluding that typing patterns contain unique characteristics that can be identified. Researchers have explained the psychological basis behind the use of keystroke dynamics so as to provide researchers with a basic understanding of the various processes involved in typing [1-5]. Extensive studies show that a person's typing pattern is a behavioral characteristic that develops over a period of time and therefore cannot be shared, lost or forgotten. The patterns show sufficiently distinct information that can be used for identification and authentication [3]. Keystroke dynamics is still an on-going research topic as evidenced from recent publications [15-17]. Researchers have proposed many methods for keystroke dynamics [3-13]. The first research paper on keystroke dynamics [3] was published by Rand Corporation in 1980, which verifies that professional typists have distinguishable "styles" of typing as measured by patterns of expected times to type certain digraphs. In [4], Young and Hammon conducted an experiment to build a template from an individual's typing manner. Monrose and Rubin [5] constructed an identification system based on template matching and Bayesian likelihood models. Hu et al. [6] proposed a K-nearest neighbor based authentication method, which focuses on improving the efficiency while maintaining the performance as other methods. In [7], the researchers presented a method based on Hidden Markov Model, which achieves a reasonable performance. Some researchers applied neural networks to keystroke dynamics [8, 9]. In [10], the authors developed a pressure-based user authentication system, where the discrete time signal is transformed into the frequency domain by using FFT. In [11-13], SVM was studied for keyboard dynamics, whose performance and efficiency are shown better than those with neural networks.

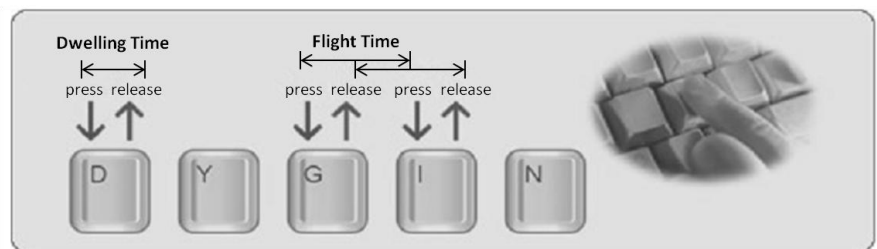

Fig. 1. The dwelling time and flight time of keystroke dynamics.

Similar to other biometrics techniques, public databases are very important to researches in the field of keystroke dynamics. However, to the best of our knowledge, there is no such database collected from a real commercialized system, while the benchmark databases in other biometric problems are generally ample. For example, the FERET and FRGC databases are commonly used in face recognition [18, 19]. In palm print recognition, the PolyU dataset [14] is the open platform for comparative evaluation of different algorithms.

The aim of a public database is to allow different researchers to test their own algorithms based on the same dataset. In the same experimental environment, the comparisons between different algorithms are more reasonable. In [3], the work on the keystroke dynamics was based on long texts. Several people were asked to type the same paragraph of words. These experiments can show the uniqueness of keystroke behavior; however, they could not be used for practical applications. In this paper, we create two available databases to help researchers interested in keystroke dynamics so that better comparisons of methods and results can be performed. It should be noted that most of the previous experiments collected test samples in laboratory environments [13]. Therefore those datasets cannot represent a more general situation. So creating an open and comprehensive database from a commercialized system for keystroke dynamics is urgently needed. Meanwhile, password is predominantly used to protect people's accounts in many applications, and so an embedded system is in great demand for real applications. In this paper, we develop an electronic device which can be embedded with the password protection technique in the system.

Previous works show that the classifier design is one of main topics in keystroke dynamics, and the one-class SVM is still the state-of-the-art algorithm [29]. In this paper, we focus more on feature extraction, and specifically investigate FFT, DCT and Gabor wavelet features to enhance the performance of the one-class SVM classifier.

The rest of the paper is organized as follows. In Section 2, we describe the proposed keystroke dynamics system and the framework of the embedded system used to create our databases, called BeiHang Keystroke Dynamics Databases. The details of the benchmark algorithms, the databases and experiments are presented in Sections 3 and 4, respectively. Section 5 gives our experiment results and Section 6 concludes this paper with some discussions on future work.

## 2. The Proposed Keystroke Dynamics Systems

In this section, two keystroke dynamics systems are presented. One is designed for the Internet, which has already been commercilized in China. The details can be found from [26-28]. Another is an Embedded system, in which we build a single electronic device for keystroke dynamics authentification. For the research purpose in the field, four datasets are collected from the two systems.

### 2.1. The Keystroke Dynamics System for the Internet

There have been several methods used to capture keystroke information. The kernel (PC) based method is powerful and can obtain any information typed on a keyboard as it reaches the operating system. However, the synchronization problem is not well solved on multi-kernel computers, though it is effective and difficult to be detected by user-mode applications on single-kernel computers. Another widely used method is based on the HOOK keyboard APIs. It includes a series of functions which reveal the status of key press or release event. However, the HOOK function is generally based on a lot of APIs, which can lead to an increase in CPU usage. There are also other methods based on web browsing, but they are not secure in keystroke event detection. To deal with the above problems, we design an instance stream to capture the key press and release events. The proposed method is effective since the instance stream is complementary to the traditional HOOK function. The whole system consists of three parts: kernel level keyboard driver, preprocessing and authenticator module(Fig. 2). It uses the keyboard driver to collect user keyboard input information. The keyboard driver can efficiently capture keystroke dynamics information, and guarantee that the keystroke event is fully recorded. In our tests, the minimum response time of the keyboard driver is just 1/1000000 second, and the system achieves an excellent performance.

A commercialized system implementing the above method was deployed to different environments, such as Internet Cafe and laboratories. It involved a variety of individuals whose registration and log-in keystroke information was collected. Each user was asked to type his/her username once and his/her password 4 or 5 times in order to create a new account. Some false data from the users' misuse of the system were included in the primary dataset. By using a filter, these false data were discarded and finally we created the BeiHang Keystroke Dynamics Database 1.

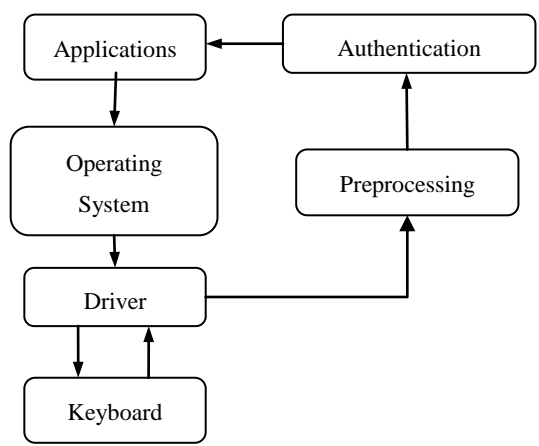

Fig. 2. Flowchart of the keystroke dynamic system.

## 2.2. The Keystroke Dynamic Device for an Embedded System

An embedded system is designed for specific functions within a larger system, often with requirement of real-time computing. It is embedded as part of a complete device usually including hardware and mechanical parts. By contrast, a general-purpose computer, such as a personal computer (PC), is designed to be flexible and to meet a wide range of end-user needs. A processor is an important unit in the embedded system hardware. It is the heart of the system. The key characteristic, however, is dedicated to handle a particular task. Since the embedded system is dedicated to specific tasks, design engineers can optimize it to reduce the size and cost of the product and increase the reliability and performance. Some embedded systems are mass-produced, benefiting from economies of scale. Complexity varies from low, with a single microcontroller chip, to very high with multiple units, peripherals and networks mounted inside a large chassis or enclosure.

The main processor of our embedded system is STC90C58AD type single-chip, which has 4kB of SDRAM and is fully competent for our task. Another main functional device is the keyboard. Generally, the keyboard used in a Single-Chip System (SCM) is specialized. These keyboards are costly, complicated and unreliable. Since the PS/2 keyboard widely used on PCs has several advantages such as low cost, compact and reliable, it is a perfect choice to use it in our SCM system. Currently, the mini-DIN 6pin is the PS/2 interface used widely on the PCs, as shown in Fig. 3. PS/2 can fulfill the slave device to the host and the host to the slave device communication. When the slave device sends data to the host device, it first checks the clock to confirm whether the clock is the high level. If so, the slave device transmits data. Otherwise, the slave device has to wait until receiving the bus right of control. When the communication finishes from the slave device to the host device, the slave device changes the data state at the time that the clock is in high level and the host device reads the data state at the moment when the clock is on its failing edge.

The PS/2 keyboard employs the second scan code set, which has two different types: make code and break code. When a key is pressed or being pressed persistently, the keyboard sends make code to the host device, and at the time that the key is released, it sends the break code of the key to the host device. The illustration is shown in Fig. 4.

With the embedded system as shown in Fig. 5, we obtained our BeiHang Keystroke Dynamics Database 2. The total cost of the device is only $15.

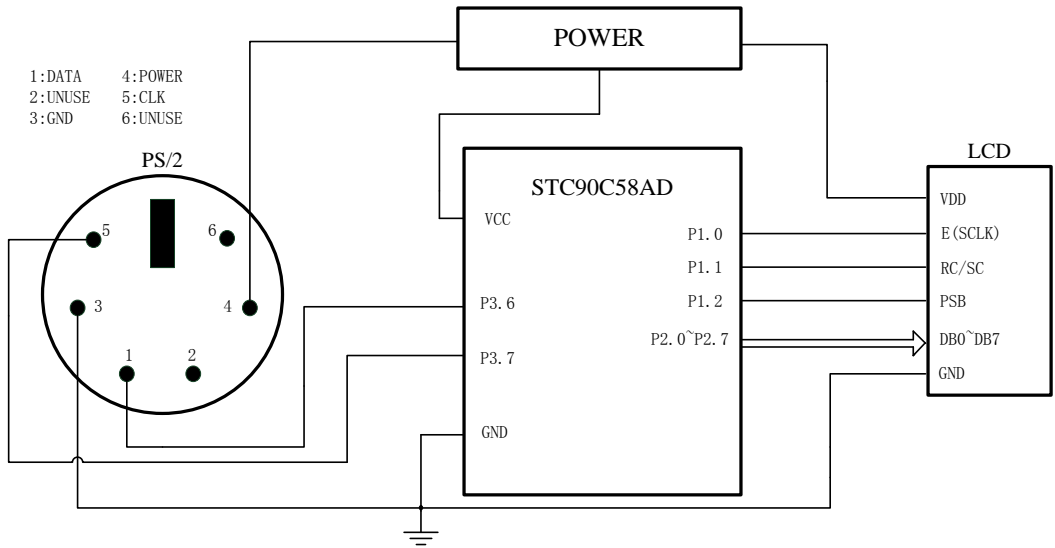

Fig. 3. The embedded device.

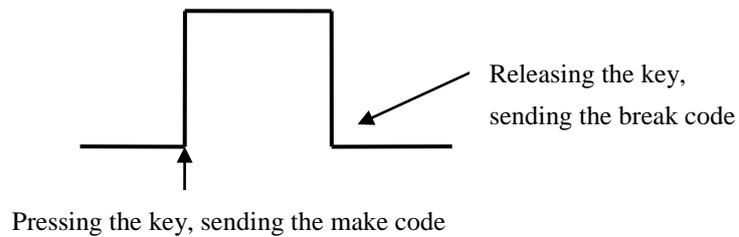

Pressing the key, sending the make code

Fig. 4 The make code and break code of the PS/2 keyboard.

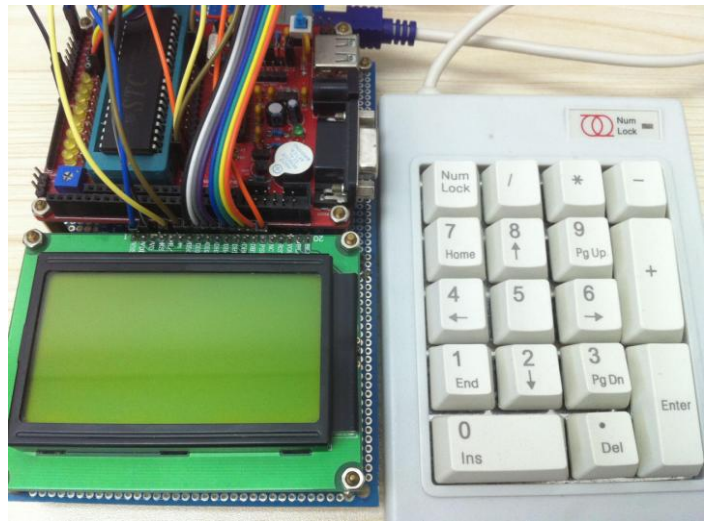

Fig. 5. The illustration of the keystroke dynamics embedded device.

## 3. The BeiHang Keystroke Dynamics Databases

We have collected two databases by using the proposed embedded system and the online system respectively. It can be used by researchers to test their algorithms and eventually boost the development of keystroke dynamics.

**3.1. Description of the Databases**

There are 209 subjects involved in building the databases. It should be noted that 10 subjects of Dataset A of Database 2 are from Dataset B of Database 2. The first database, named BeiHang Keystroke Dynamics Database 1, is captured by the online system, and the second one, named BeiHang Keystroke Dynamics Database 2, is collected from the embedded system. The subjects gather registration data from genuine users used as training samples, log-in data from genuine users and log-in data from intruders. All the data are stored in text format; they can be downloaded at [26].

In each folder of Database 1, the only training file contains 4 or 5 registration samples and the file name is in the format of, say, []12345(-regliaoxiaoying).txt meaning that this is the training file for ID being 12345 and password being liaoxiaoying with [] being the label of the file. All the testing files have the same format: [Year-Month-Day Hour.Min.Sec]ID(-loginPSW)_ IsGenuine_IsPostive.txt, where IsGenuine = 0 or 1 represents the data from a genuine user or an intruders; IsPostive = y or n represents the positive data from a user or the negative data from an intruder. For example, the testing file, [2009-12-30 14.07.01]12345(-loginliaoxiaoying)_1_n.txt, indicates that the login time is 2009-12-30 14.07.01, ID is 12345, PSW is liaoxiaoying, and it is negative data from an intruder.

The file names in Database 2 have been simplified. The folders are named as PSW or the time when the data were collected. In the folder, [].txt stores genuine user registration data. The entire testing files are in the form of time-index_IsGenuine_ IsPostive.txt.

The BeiHang Keystroke Dynamics Database 1 includes 1902 test samples and 477 training samples from 117 subjects. The whole Database 1 is divided into two subsets, Dataset A and Dataset B, collected from two different environments. Dataset A was collected in Internet Cafe. It contains 49 subjects, 212 training samples, 157 testing samples from genuine users and 996 testing samples from intruders, as shown in Table 1. The developed commercial system was embedded into the login system of an online application. In Database 1, Dataset B was collected online in a university lab. It contains 68 subjects, 265 training samples, 214 testing samples from genuine users and 535 testing samples from intruders. The BeiHang Keystroke Dynamics Database 2 was collected by the embedded system, which contains 5089 test samples and 478 training samples from 92 subjects. Dataset A and B in Database 2 are released for research purpose. Dataset A of Database 2 contains 52 subjects, 228 training samples, 717 testing samples from genuine users and 1468 testing samples from intruders. Dataset B of database 2 contains 50 subjects, 250 training samples, 1103 testing samples from genuine users and 1801 testing samples from intruders. The details are given in Table 1. It is worth noting that there are 10 subjects appear both in these two subsets. which contain data of stable typing rhythms.

All the data in these databases are originally collected, without any manual modification. Generally a password is represented by the following stream: $P_1, R_1, P_2, R_2, ..., P_n, R_n$, where $P_i$ and $R_i$ represent the press and release time of the $i$th keystroke of a password. The meanings of different files are shown by their file names.

Table 1. Composition of the databases.

|  |  | Database 1 | | Database 2 | |
| --- | --- | --- | --- | --- | --- |
|  |  | Dataset A | Dataset B | Dataset A | Dataset B |
| Number of subjects | | 49 | 68 | 52 | 50 |
| Number of training samples | | 212 | 265 | 228 | 250 |
| Number of testing samples | genuine users | 157 | 214 | 717 | 1103 |
| | intruders | 996 | 535 | 1468 | 1801 |

### 3.2. Database Access

To download the databases for research purpose, one can visit mpl.buaa.edu.cn, or send an email to the corresponding author.

## 4. Benchmark Algorithms

The framework of our Keystroke Dynamic System is shown in Fig. 6. Feature extraction and classification algorithm are the main components and are discussed in detail in the following sections.

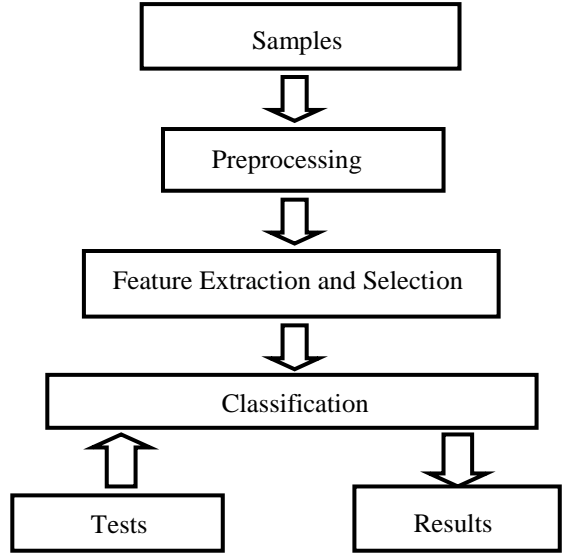

Fig. 6. Framework of the Keystroke Dynamics System.

**4.1. Feature Extractions**

Suppose a password is represented by the following sequence:

$$P_1, R_1, P_2, R_2, ..., P_n, R_n \qquad (1)$$

where $P_i$ and $R_i$ represent the press and release time of the $i$th keystroke of a password. The elements of the feature vector extracted from the original keystroke information are classified into two categories: dwelling time and flight time. The dwelling time is calculated by $R_i - P_i$, and the flight time by $P_i - R_{i-1}$. Therefore, the extracted feature from the original sequence is represented as:

$$I = (R_1 - P_1, P_2 - R_1, R_2 - P_2, ..., P_n - R_{n-1}, R_n - P_n). \qquad (2)$$

The above feature is also called the original feature. The number of the registration samples collected in the training procedure is 4 or 5..

In signal processing and object recognition, the transformed features are extensively studied [22-25], such as Fast Fourier Transform (FFT), Discrete Cosine Transform (DCT), and Gabor wavelets. This paper investigates these three transformation methods to further enhance the performance of keystroke dynamics systems using the original feature directly. Different from most of previous works focusing on classifier design, this paper design better features for performance improvement.

The Fast Fourier Transform (FFT) is first discussed by Cooley and Tukey [25]. FFT, as an important tool in the frequency domain analysis, is widely used in the field of signal processing. A discrete Fourier transform can be computed using an FFT, which is a discrete Fourier transform algorithm and it reduces the number of computations needed for $N$-sized signal from $2N^2$ to $2N \log N$. In implementation, we directly use the FFT() function provided in Matlab software. The FFT feature is represented as $I_F$, length of which can be any times to original input signal by controlling the parameters.

A Discrete Cosine Transform (DCT) expresses a finite sequence of data points in terms of a sum of cosine functions oscillating at different frequencies [24]. DCTs are important to numerous applications from signal compression to object representation. A DCT is a Fourier-related transform similar to the Discrete Fourier Transform (DFT), but using only real numbers. In this paper, we directly use the DCT() function provided in Matlab software. The DCT feature is represented as $I_D$, length of which can be any times to original input signal by controlling the parameters, too.

FFT and DCT are two commonly used transforms for signal analysis. Taking FFT and DCT of $I(n)$, we have the FFT and DCT features represented as $I_F(n)$ and $I_D(n)$, respectively.

Gabor features are widely used as feature descriptors extracted by a set of Gabor wavelets (kernels) which model the receptive field profiles of cortical simple cells [22,23]. They can capture the salient visual properties in the input signal, such as spatial characteristics, because the kernels can selectively enhance the features in certain scales and orientations. Here, we obtain Gabor feature from the original keystroke dynamics feature to enhance the object the representation capability. The 2D Gabor wavelets (kernels, filters) can be defined as

$$G(x,y) = \frac{1}{2\pi s_x s_y} \exp(-\frac{1}{2}[(\frac{x}{s_x})^2 + (\frac{y}{s_y})^2] + 2\pi i(Ux+Vy)]) \qquad (3)$$

where $i = \sqrt{-1}$; the Gabor filter is basically a Gaussian (with variances $s_x$ and $s_y$ along the $x$- and $y$-axes respectively) modulated by a complex sinusoid (with centre frequencies $U$ and $V$ along $x$- and $y$-axes respectively). Sicne the original feature is 1D in our application, we use 1D Gabor filter as

$$G(n) = \frac{1}{2\pi s_n} \exp\{-\frac{1}{2}(\frac{n}{s_n})^2 + 2\pi i U n\} \tag{4}$$

where $n$ is of the index of the input feature and $s_n$ is the maximum variance. The Gabor feature is then obtained by the convolution operation

$$I_g(n) = I(n) * G(n). \tag{5}$$

Let the magnitude part of $I_g(n)$ be $I_{g,M}^{(n)}$. From $m$ different $U$s, we can get $m$ groups of the Gabor features, which are put together to form a feature matrix of size $1 \times nm$:

$$I_G = (I_{g,M}^{(1)}, I_{g,M}^{(2)}, ..., I_{g,M}^{(m)}) \tag{6}$$

**4.2. Classification Algorithm**

Support Vector Machine (SVM) is a typical classifier in machine learning with top perforamnces in many applications. One Class SVM (OC-SVM) as a variant of SVM can train a classification model only from one class without negative samples. OC-SVM can also be viewed as a regular two-class SVM where all the training data lie in the first class. The keystroke dynamics is basically a single-class problem. In this paper we exploit the nonlinear version of the OC-SVM algorithm which maps the input data into a high dimensional feature space (via a kernel function) as our first classification algorithm. It should be noted that OC-SVM is also one of the state-of-art methods in keystroke dynamics.

SVM for one-class classification has been widely investigated. Researchers proposed a method of adapting the SVM methodology to the one-class classification problem in [20,21]. Essentially, after transforming the features via a kernel, they treat the origin as the only member of the second class. By using relaxation parameters, they can separate the image of the one class from the origin. Then the standard two-class SVM techniques are employed. As shown in Eq.(7), they phrased the problem in the following way: Suppose that a dataset has a probability distribution P in the feature space. Find a "simple" subset S of the feature space such that the probability that a test point from P lies outside S is bounded by some a priori specified value. It can be summarized as mapping the data into a feature space using an appropriate kernel function, and then trying to separate the mapped vectors from the origin with maximum margin (see Fig. 7).

$$\begin{aligned} \text{Min} \quad & \frac{1}{2}\|w\|^2 + \frac{1}{vl}\sum_{i=1}^{l}\xi_i - \rho \\ s.t. \quad & (w \cdot \Phi(x_i)) >= \rho - \xi_i \ \ i=1,2,...l \ \ \xi_i >= 0 \end{aligned} \tag{7}$$

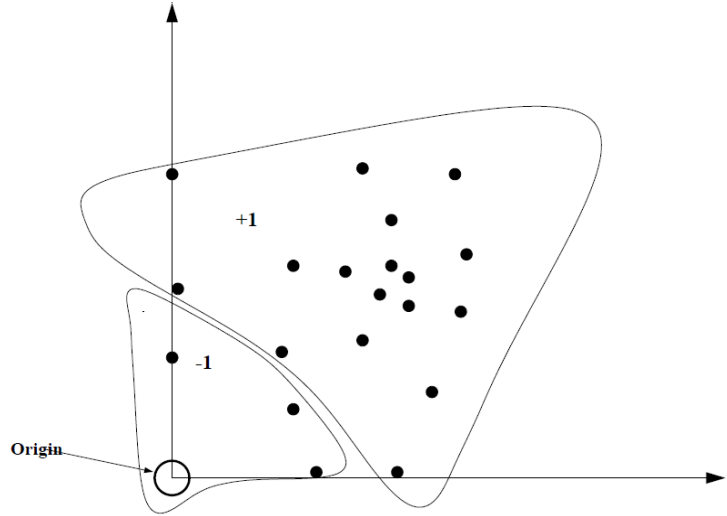

Fig. 7. One-Class SVM.

Among the various supervised methods for statistical pattern recognition, the Nearest Neighbor (NN) rule achieves consistently good performance, without a priori assumptions about the distributions from which the training examples are drawn. It involves a training set of both positive and negative samples. A new sample is classified by calculating the distance to the nearest training sample; the label of the training sample then determines the classification of the new sample. We use the NN classifier as the baseline algorithm on the collected database. In the NN classifier, the Euclidean is tested as the measure with given threshold.

$$Dis_X(x') = \min_i \quad \|x' - x_i\|, x_i \in X, i = 1, 2, ..., n \tag{8}$$

where $x'$, $x_i$ are input vectors.

Gaussian classification is another common methodology in object recognition. In a Gaussian model, a set of data is characterized by the mean and covariance for each class within the data along the dimensions. New data points can then be classified by measuring their distances from each centroid and assigning the class of the closest centroid to it. In probability theory, the Gaussian distribution, is a continuous probability distribution that is often used as a first approximation to describe real-valued random variables that tend to cluster around a single mean value [1]. Suppose that the keystroke data follow the Gaussian distribution, and therefore we can build a Gaussian classifier using the training data. Then we can use the model for authentication. The Gaussian distribution is

$$p(x) = \frac{1}{2\pi^{d/2}|\Sigma|^{1/2}} \exp\{-\frac{1}{2}(x-\mu)^T \Sigma^{-1}(x-\mu)\} \tag{9}$$

where $x$ is a test sample and $\Sigma$ and $\mu$ are the covariance matrix and mean vector of the Gaussian model respectively. Noted that $\Sigma$ is assumed to be a diagonal matrix to enhance computation efficiency.

### 4.3. Fusion of Features and Classifiers

Combination of different expert decisions is extensively studied in previous twenty years. Combination methods can be grouped by the level at which they operate. The simplest way is in the feature level, where different kinds of features are concatenated into an extended feature vector. This combination inherits the advantages of different features, and any classifier is easily used with them to build the final classification model. Combination can also be done in the level of decision or output score, which is called classifier-level combination. It is a quite popular way as the score is generally considered as a new kind of feature. This paper investigates both methods for performance improvement. For feature-level combination, we can easily get the new extended feature as

$$I_{feature-level} = [I, I_G, I_F, I_D] \qquad (10)$$

where $I$ and $I_G$ are obtained by Equations (2) and (6), $I_F$ and $I_D$ are the FFT and DCT features, respectively. Similar to the feature-level combination, the classifier-level combination is based on the scores of classifiers. Supposed we have $k$ classifiers, whose scores are denoted as $score_i, i = 1, 2, \cdots k$. The combination can be formulated as

$$I_{classifier-level} = [score_1, score_2, ..., score_k]. \qquad (11)$$

### 5. Experiments

In this section, we present benchmark experimental results for some classification and feature extraction algorithms on the BeiHang Keystroke Dynamics Databases. The extensive experiments include the evaluation of different features, classifiers, and their fusions. We also show that specific rhythms for different individuals can lead to high performance, which can be used in practical applications, such as password protection.

### 5.1. Evaluation Criteria

In the experiments, we use the False Positive Rate and the True Positive Rate for evaluation metrics. The former is the percentage of intruders who can enter the accounts by imitating the typing manner of genuine users. The latter is the percentage of genuine users who can successfully log into the system with the right keystroke manner. By changing the threshold in the classification procedure, we obtain a series of True Positive Rates and False Positive Rates, and then we use these results to draw a ROC curve. The ROC curve is used for evaluation of various algorithms including the OC-SVM classifier, Gaussian classifier, and NN classifier with the original feature, Gabor, DCT, and FFT features, etc.. We also provide the Equal Error Rate (EER) for further evaluation of different methods. EER is the percentage where the False Positive Rate equals the False Negative Rate.

**5.2. Experimental Results**

A. *Comparative experiments in terms of the ROC curves*

(1) The ROC curves with the original feature and OC-SVM on different datasets

The comparative experiments on different datasets are first conducted with the original feature defined in Equation (2) and the OC-SVM classifier. OC-SVM uses the RBF kernel function, and the algorithm is form LIBSVM [21]. In the experiments, the training set is used in registration process, while the test set is for performance evaluation. We first use the original feature to get the baseline results. Dataset 1A in Fig. 8 represents Dataset A of Database 1. Similar representations are foe Database 2B, Database 2A and Database 2B. With the original feature and the OC-SVM classifier, the experimental results in Fig. 8 demonstrate that the performances on Dataset A of Database 1 and Dataset B of Database 2 are better than those on the other two datasets. Dataset A of Database 2 is a very difficult one, which needs better features and classifiers for satisfactory results. The excellent result can be achieved on Dataset B of Database 2, which provides a suitable way to use keystroke dynamics in real applications, by choosing a stable typing habit in the keystroke process.

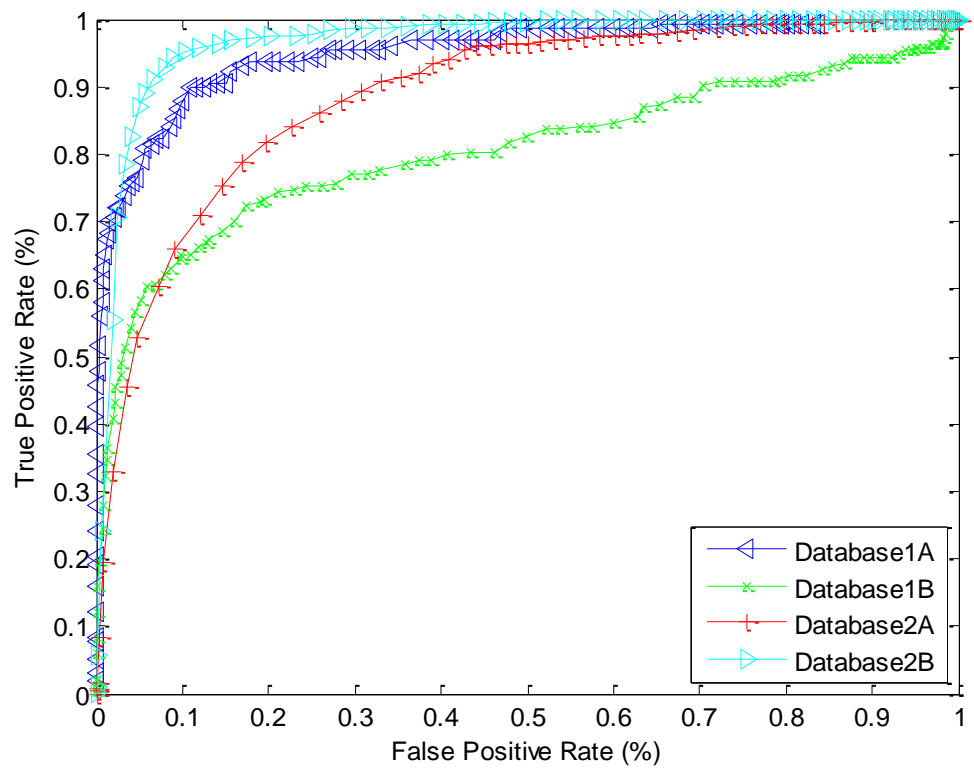

Fig. 8. Comparison of the ROC results of the four datasets.

(2) Experiments with different features and the OC-SVM classifier

Extensive experiments are performed to evaluate the effectiveness of different features. Since Database 1 has been well discussed in our previous work [29], all the ROC curves shown here are mainly conducted on BeiHang Keystroke Dynamics Database 2. As shown in Figs. 9 and 10, the features for genuine users and intruders are obviously different. The samples of the genuine user and intruders are from folder 8_8_560727 in Dataset A of Database 2. The length of the original feature is 11, when the length of the password is 6. The lengths of other features are determined by their own parameters. The Gabor feature is twice as long as the original one and the FFT and DCT features quadruple. Some samples are visualized in Fig. 9 and Fig. 10.

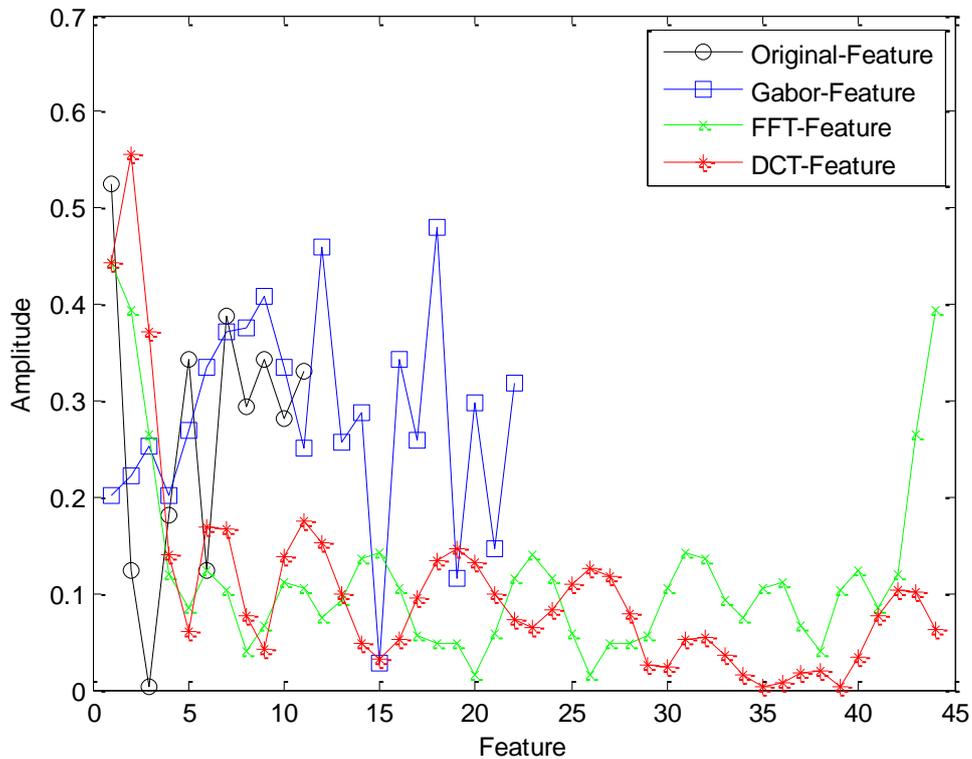

Fig. 9 The visualization of the original, FFT, DCT and Gabor features for genuine user samples in 8_8_560727.

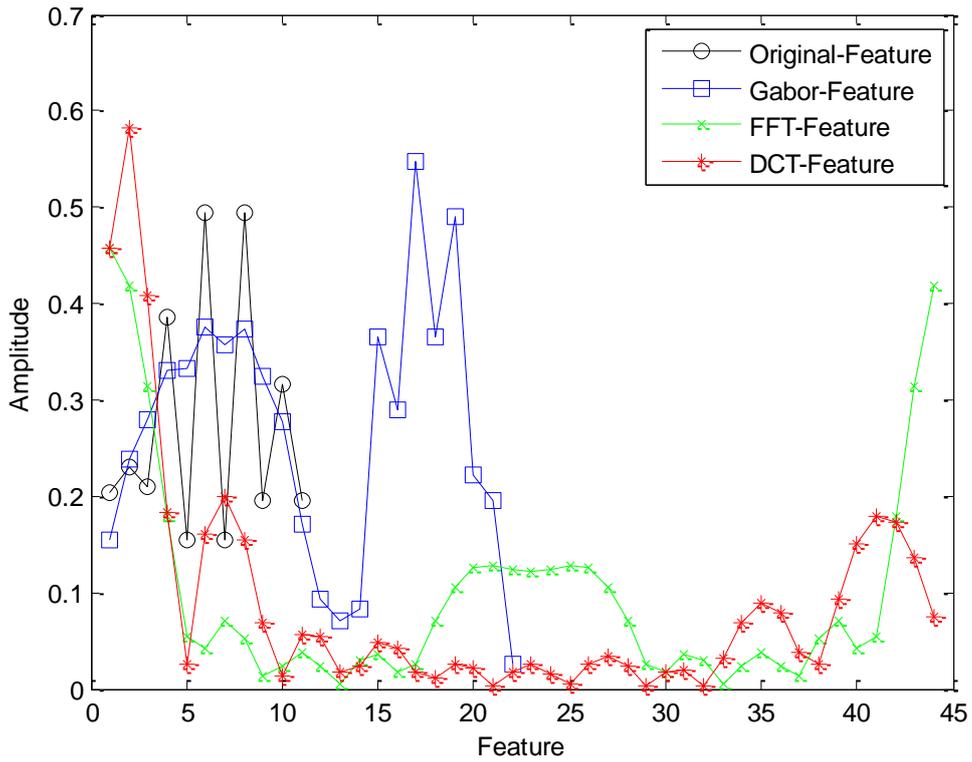

Fig. 10. The visualization of the original, FFT, DCT and Gabor features for intruder samples in 8_8_560727.

To see what extent the Gabor, FFT and DCT features affect the performance, we conduct the experiments to observe performance variations. Fig. 11 shows that the performance is significantly improved with the Gabor feature on Dataset A of database 2 in terms of the ROC curve, while the FFT and DCT features are slightly worse than the original feature.

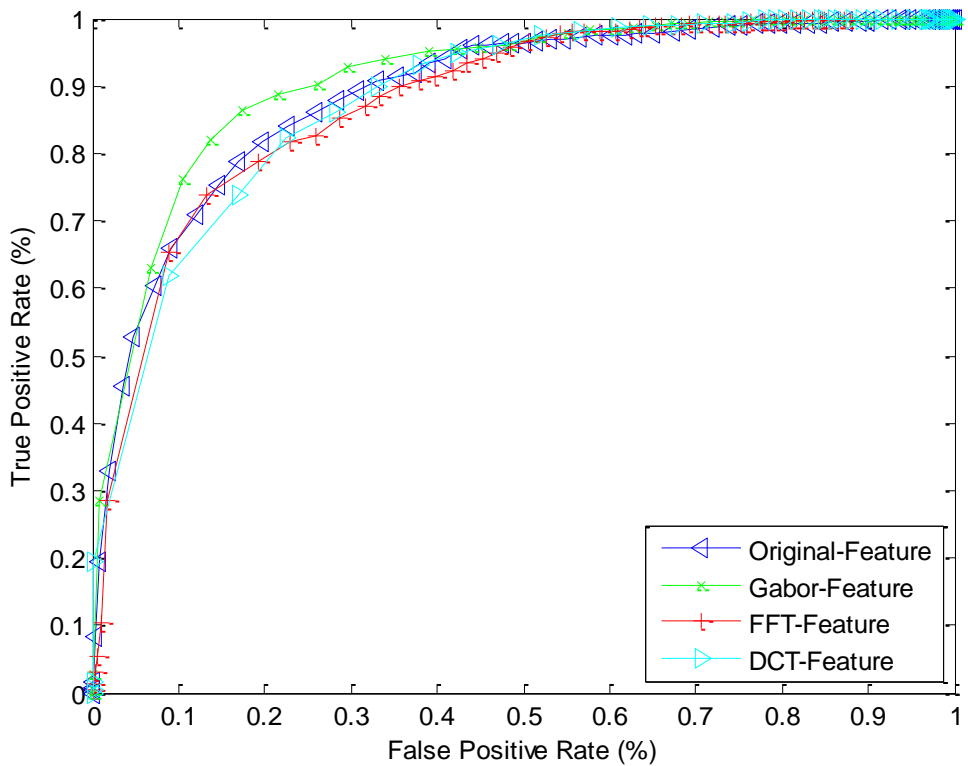

Fig. 11 Experimental results of of the four features on Dataset A of Database 2 with OC-SVM.

We also test the performance of the simple combination in the feature level, i.e., concatenating the original feature with Gabor, FFT or DCT feature. Three extended features are generated, called OriGabor, OriFFT and OriDCT. The comparison is shown in Fig. 12 where these three new combined features all outperform the original feature on Dataset A of Database 2.

The improvements can also be observed from Fig. 13 and Fig. 14. The curve of the OriGabor feature rises faster nearby the origin. The Gabor feature and the OriGabor feature perform much better than the original one. It can also be observed that the OriFFT feature achieves a better performance than the original feature, even when the single FFT feature is worse than the original one.

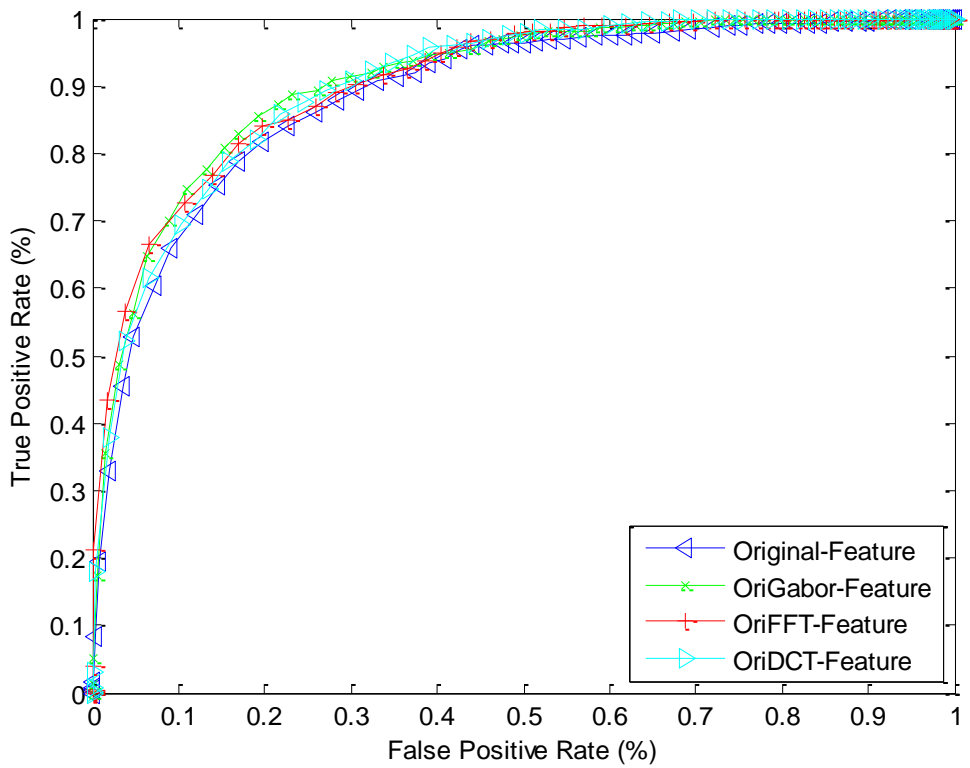

Fig. 12. Experimental results of four features on Dataset A of Database 2 with OC-SVM.

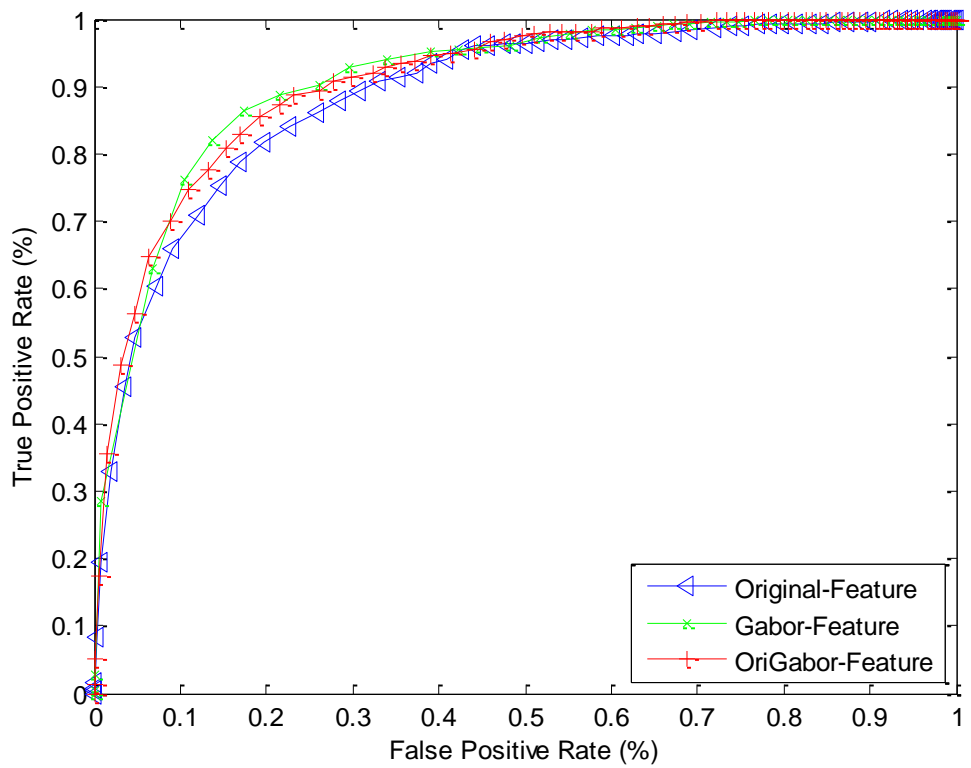

Fig. 13 Experimental results of three features on Dataset A of Database 2 with OC-SVM.

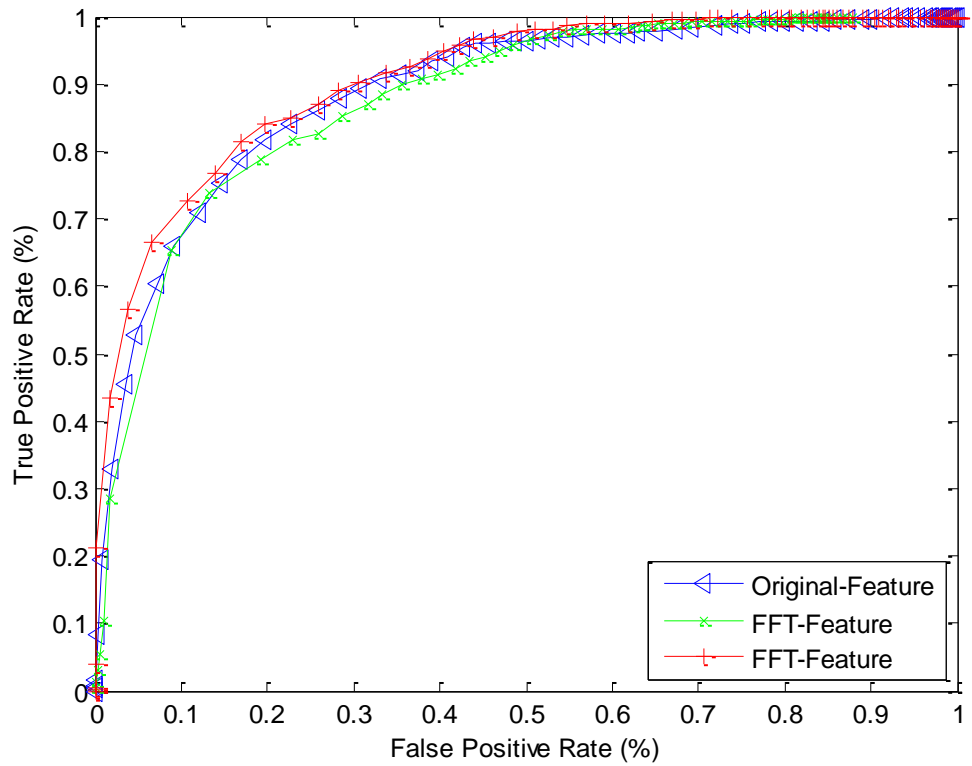

Fig. 14 Experimental results of three features on Dataset A of Database 2 with OC-SVM

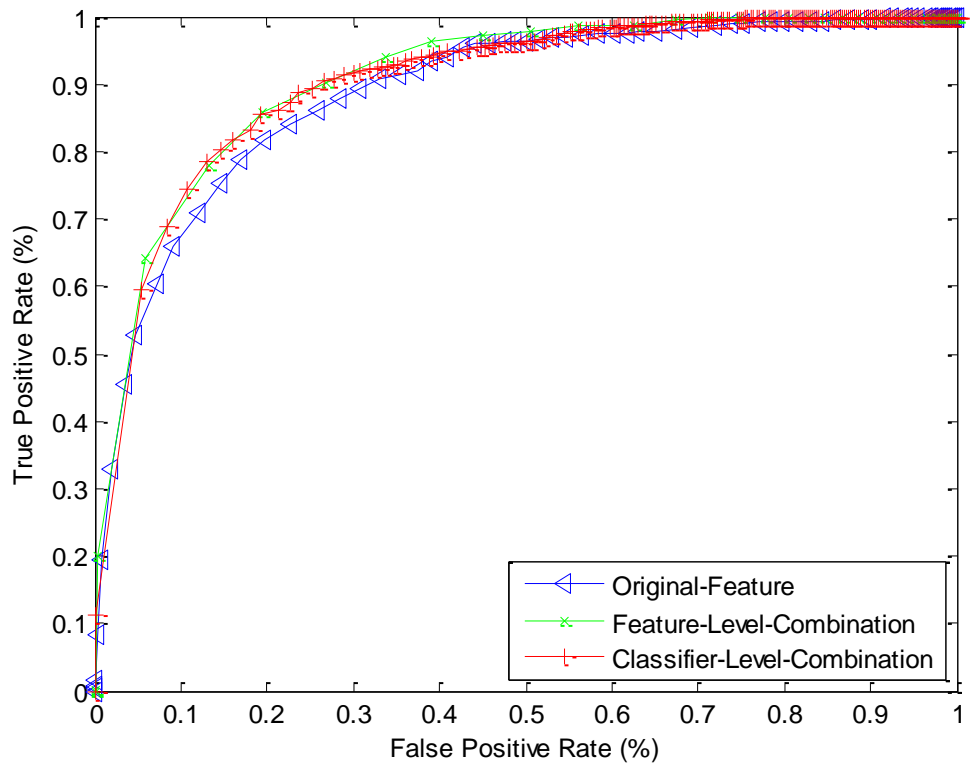

Fig. 15. Experimental results of the feature-level and classifier-level combination compared with the original feature on Dataset A of Database 2 with OC-SVM

(3) Experiments on feature-level and classifier-level combination

These experiments are also designed to evaluate the effectiveness of feature-level and classifier-level combination. $I_{feature-level}$ in Equation (11) and $I_{classifier-level}$ in Equation (12) are applied in classification. Fig. 15 illustrates that the classifier-level combination performs better than the original feature. These results reveal that the more features are extracted from the raw input signal, the more effective information is reserved.

(4) Comparative experiments with customization of keystroke rhythm

In this section, we perform an experiment on different datasets to show how the specific rhythm of keystroke dynamics affects the final performance, which provides a feasible way for commercial applications. On Dataset B of Database 2, we customize the rhythm of keystroke, that is to say, the dwelling and flight times are different for different input characters, which can be seen as a specific feature for different individuals. As shown in Fig. 16 and Fig.17, the performance on Dataset B of Database 2 is significantly better than those on Dataset A, which is not customized on the keystroke process.

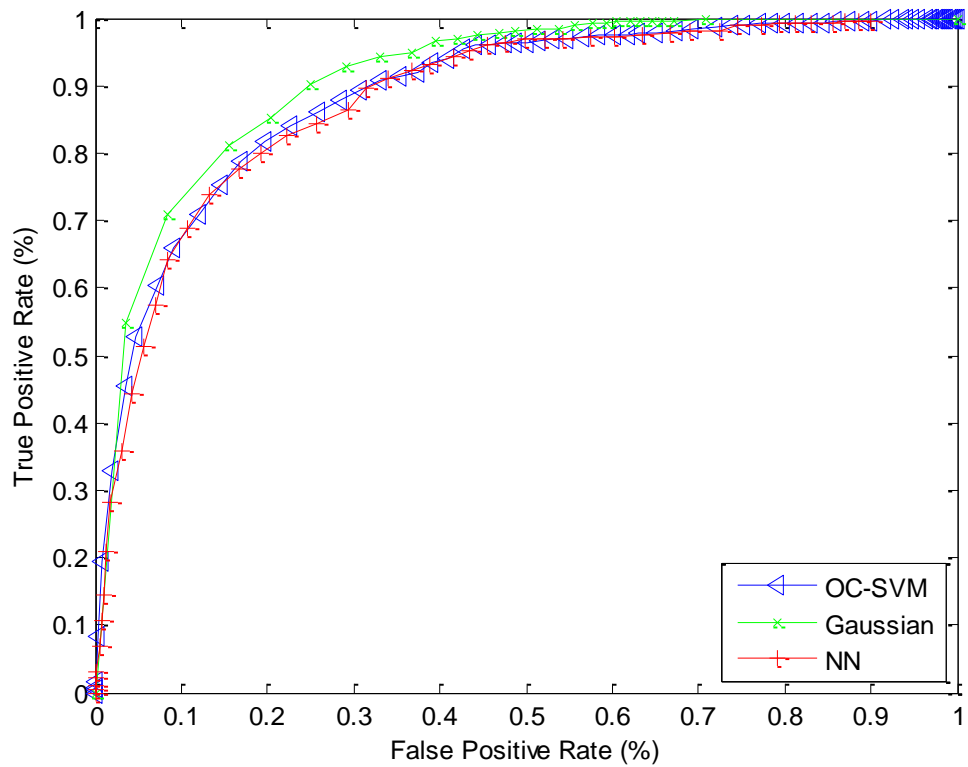

Fig. 16 Experimental results of different classifier on Dataset A of Database 2 with the original feature.

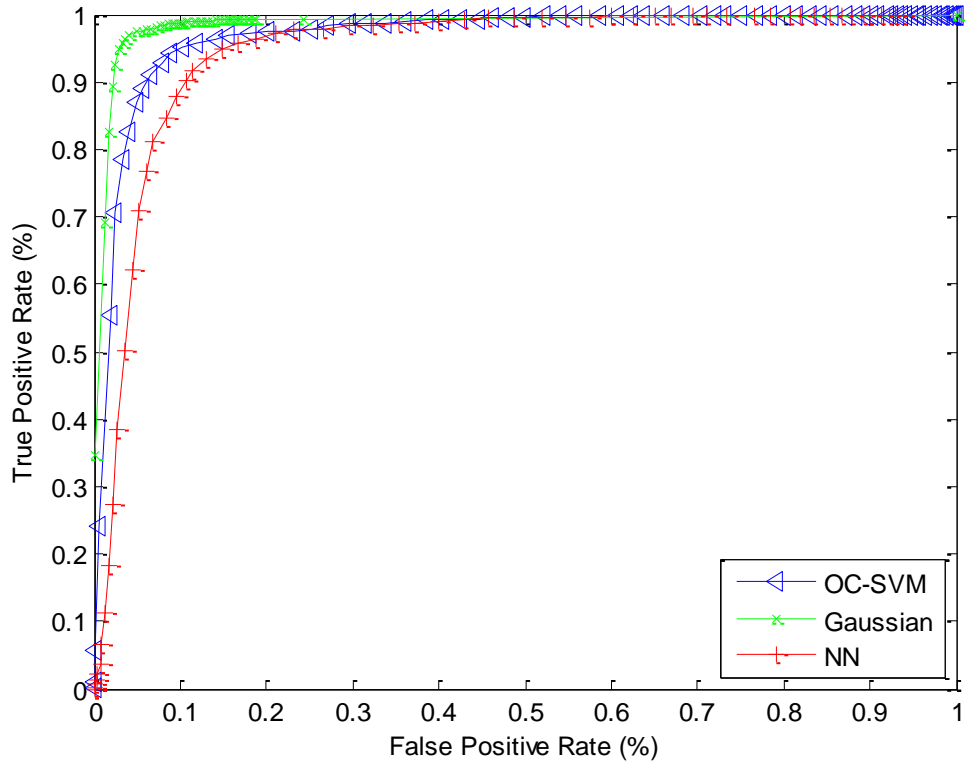

Fig. 17 Experimental results of different classifier on Dataset B of Database 2 with the original feature

B. *Comparative experiments in terms of the Equal Error Rate (EER)*

To further compare the performance of different features and classifiers, we show the experimental results in terms of Equal Error Rate (EER) in Tables 2-5. Part of the results contains those from our previous paper[29]. The experimental results in Table.2 show that the Gaussian classifier with the original feature achieves much better performance than the OC-SVM and NN classifiers, while the NN classifier is worst. It is indicated in Table 3 that on Dataset B of Database 2, the Gaussian classifier greatly outperforms the other classifiers. However, the results on Datasets A and B of database 1 as shown in Tables 4-5 demonstrate that OC-SVM is better than the Gaussian classifier.

Table 2. EER results with different features and classifiers on Dataset A of Database 2.

| Dataset A of database 2 | OC-SVM | Gaussian | NN |
|---|---|---|---|
| Original-Feature | 19.044 | 17.350 | 19.695 |
| Gabor-Feature | 15.649 | 17.556 | 16.437 |
| FFT-Feature | 20.308 | 19.176 | 19.134 |
| DCT-Feature | 20.522 | 17.668 | 19.806 |
| OriGabor-Feature | 16.980 | 16.734 | 16.437 |
| OriFFT-Feature | 17.758 | 17.001 | 17.724 |
| OriDCT-Feature | 18.458 | 16.294 | 18.262 |
| Feature-level-combination | 17.142 | | |
| Classifier-level-combination | 17.065 | | |

Table 3. EER results with different features and classifiers on Dataset B of Database 2

| Dataset B of database 2 | OC-SVM | Gaussian | NN |
|---|---|---|---|
| Original-Feature | 7.6642 | 3.6846 | 10.36 |
| Gabor-Feature | 4.6143 | 4.9463 | 5.854 |
| FFT-Feature | 8.1224 | 7.2872 | 9.2172 |
| DCT-Feature | 6.7049 | 4.4077 | 9.9336 |
| OriGabor-Feature | 4.6729 | 4.1365 | 5.2063 |
| OriFFT-Feature | 5.1547 | 4.6488 | 6.8225 |
| OriDCT-Feature | 5.0363 | 4.2754 | 8.3795 |
| Feature-level-combination | 4.6078 | | |
| Classifier-level-combination | 5.6626 | | |

Table. 4 EER results with different features and classifiers on Dataset A of Database 1, where * indicates results from [29].

| Dataset A of database 1 | OC-SVM | Gaussian |
|---|---|---|
| Original-Feature | 12.1886* | 14.1459* |
| Gabor-Feature | 14.501 | 20.143 |
| FFT-Feature | 19.391 | 22.755 |
| DCT-Feature | 20.382 | 23.414 |
| OriGabor-Feature | 11.997 | 19.745 |
| OriFFT-Feature | 12.674 | 20.998 |
| OriDCT-Feature | 12.577 | 21.019 |
| Feature-level-combination | 14.291 | |
| Classifier-level-combination | 12.058 | |

Table 5 EER results with different features and classifiers on Dataset B of Database 1, where * indicates results from [29].

| Dataset B of database 1 | OC-SVM | Gaussian |
|---|---|---|
| Original-Feature | 25.3559* | 28.2028* |
| Gabor-Feature | 27.388 | 31.421 |
| FFT-Feature | 31.464 | 32.710 |
| DCT-Feature | 32.886 | 35.514 |
| OriGabor-Feature | 25.234 | 34.579 |
| OriFFT-Feature | 25.762 | 35.184 |
| OriDCT-Feature | 26.272 | 35.981 |
| Feature-level-combination | 26.954 | |
| Classifier-level-combination | 26.168 | |

## 6. Conclusions

Two large databases have been collected and open for public research. Different features and benchmark algorithms have been tested and summarized. We design both an embedded password protection device and an online keystroke dynamics system, which is the first commercialized system in China. The new feature include Gabor, FFT, DCT and their combinations. The benchmark results are obtained by the one-class support vector machine, Gaussian model, and nearest neighbour classifier, applied on the original and extended features. Our future work will focus on boosting the classifiers and promoting the applications.

**Acknowledgement**

This work was supported in part by the Natural Science Foundation of China, under Contracts 60903065, 61039003, and 61272052, in part by the Ph.D. Programs Foundation of Ministry of Education of China, under Grant 20091102120001, and in part by the Fundamental Research Funds for the Central Universities. We thank the volunteers for the database collection. Thanks Yilin Li, and He Gong, who are former members of MPL Lab of Beihang University, and also thank for Prof. Yongsheng Gao, for his comments on the paper.